\documentclass [aps,prl,twocolumn] {revtex4}
\usepackage[final]{graphics}
\usepackage{amssymb}
\usepackage{amsfonts}
\usepackage{epsfig}

\begin{document}


\hspace{15.cm} 1

\title{Monte Carlo simulations of ordering in ferromagnetic-antiferromagnetic bilayers}

\author{Shan-Ho Tsai$^{a,c}$, D. P. Landau$^a$, and Thomas C. Schulthess$^b$}
\affiliation{
$^a$ Center for Simulational Physics, University of Georgia, Athens, GA 30602\\
$^b$ Center for Computational Sciences and Computer Science \&
Mathematics Division, Oak Ridge National Laboratory, Oak Ridge TN 37831\\
$^c$ Enterprise Information Technology Services, University of Georgia, Athens, GA 30602}

\begin{abstract}
Monte Carlo simulations have been used to study phase transitions on
coupled anisotropic ferro/antiferromagnetic (FM/AFM) films
of classical Heisenberg spins. We consider films of different thicknesses,
with fully compensated exchange across the FM/AFM interface. 
We find indications of a phase transition on each film, occuring
at different temperatures. It appears that both transition temperatures
depend on the film thickness.
\end{abstract}


\maketitle

\newpage

\section{Introduction}

Most of the attention given to ferromagnetic/antiferromagnetic (FM/AFM) 
bilayers is
normally directed to the changes of the magnetic properties of the
FM when it is in contact with the AFM. The commonly observed effects are
a unidirectional shift (exchange bias) and a significant increase of the
coercivity \cite{reviews}. The blocking
temperature, $T_B$, below which these effects are observed, is comparable to
the bulk Neel temperature when the AFM film is thick, but can be
considerably lower when the AFM film is thin 
\cite{parkin,vanderzaag96,ambrose,devasahayam,vandriel}. 
It is generally believed
that this reduction of the blocking temperature is due to finite size
effects in the AFM, {\it i.e.} the ordering temperature of the AFM is
decreased due to its finite thickness \cite{ambrose}.  
In a recent neutron experiment on
CoO/Fe$_3$O$_4$, however, van der Zaag {\it et. al.} \cite{vanderzaag00} 
have found that the AFM shows signs of ordering above $T_B$. This indicates 
that the proximity of the FM influences the
phase transition in the AFM in a way that cannot be predicted from
studying free AFM films.

In this paper we use Monte Carlo simulations to analyze the
effect of the FM on the transition between disordered and ordered states
in the AFM for various film thicknesses and compensated
exchange across the FM/AFM interface.

\section{Model and Methods}
The system studied here consists of a multi-layered ferromagnetic (FM) film 
coupled to an underlying multi-layered antiferromagnetic (AFM) film with no
lattice mismatch at the FM/AFM interface. The Hamiltonian of the model is 
given by
\begin{eqnarray}
{\cal H}&=&-J_F\sum_{\langle{\bf r},{\bf r'}\rangle\in {\rm FM}}{\bf S_r}\cdot {\bf S_{r'}}
-K_F\sum_{{\bf r}\in {\rm FM}}(S_{\bf r}^z)^2 \nonumber\\  
&&-J_A\sum_{\langle{\bf r},{\bf r'}\rangle\in {\rm AFM}}{\bf S_r}\cdot {\bf S_{r'}}
-K_A\sum_{{\bf r}\in {\rm AFM}}(S_{\bf r}^y)^2 \nonumber \\ 
&&-J_{AF}\sum_{\langle{\bf r},{\bf r'}\rangle\in {\rm FM/AFM}}{\bf S_r}\cdot {\bf S_{r'}}
\end{eqnarray}
where ${\bf S_r}=(S_{\bf r}^x,S_{\bf r}^y,S_{\bf r}^z)$ is a 
three-dimensional classical Heisenberg spin of unit length, 
$\langle{\bf r},{\bf r}'\rangle$ denotes nearest-neighbor pairs of spins 
coupled with exchange interactions $J_F>0$ on the FM film, $J_A<0$ on the
AFM film, and $J_{AF}$ at the FM/AFM interface, which is on a (001) plane. 
Spins on the AFM film have a uniaxial single-site anisotropy $K_A>0$, whose
easy axis is along the $y$ axis. In contrast, the single-site anisotropy 
for spins on the FM film has a hard-axis ($K_F<0$) along the $z$ direction, 
which is perpendicular to the FM/AFM interfacial plane. No external magnetic 
field is applied on the films. 
The structure of the films is a body-centered cubic lattice,
with linear sizes $L_x$, $L_y$, and $L_z^A+L_z^F$, measured in terms of 
two-spin unit cells. $L_z^A$ and $L_z^F$ denote the number of unit-cell 
layers on the AFM and FM films, respectively.
Hence, the total number of spins in the lattice is $N=2L_xL_y(L_z^A+L_z^F)$.
We use periodic boundary conditions along the $x$ and $y$ directions and free
boundary conditions along the $z$ direction.

The effect of the film thickness on the FM and AFM transition temperatures is 
studied for fixed interaction parameters. We consider $J_F=J>0$, $J_A=-J$, 
$K_A=J$, $K_F=-0.5J$, and a compensated interface with $J_{AF}=r\:J$, 
where $r$ is a random number uniformly sampled in the interval $[-1,1]$. 
Values of the film thickness used are $L_z^A=L_z^F=3$, $6$, and $12$, 
with several cross sections 
($L_x=L_y=12,20,40$, and $60$) to analyze finite-size effects. 

The order parameter for the ferromagnetic transition is the uniform 
magnetization per spin $m=|\sum_{\bf r}{\bf S_r}|/N$, whereas to 
characterize the antiferromagnetic transition it is necessary to divide 
the BCC lattice into two simple cubic sublattices, denoted I and II, and 
consider the staggered magnetization per spin defined as 
$m_s=|\sum_{{\bf r}\in I}{\bf S_r}-\sum_{{\bf r}\in II}{\bf S_r}|/N$.

Our simulations were carried out using importance sampling Monte Carlo methods
 \cite{MCbook}, with Metropolis algorithm, at fixed temperature $T$. 
Typically $3\times 10^5$ 
Monte Carlo Steps/site (MCS) were used for computing averages after about 
$1 \times 10^5$ MCS were discarded for thermalization. Whenever not shown, 
error bars in the figures are smaller than the symbol sizes.

\section{Results}
\begin{figure}[ht]
\centering
\leavevmode
\includegraphics[clip,angle=0,width=8cm]{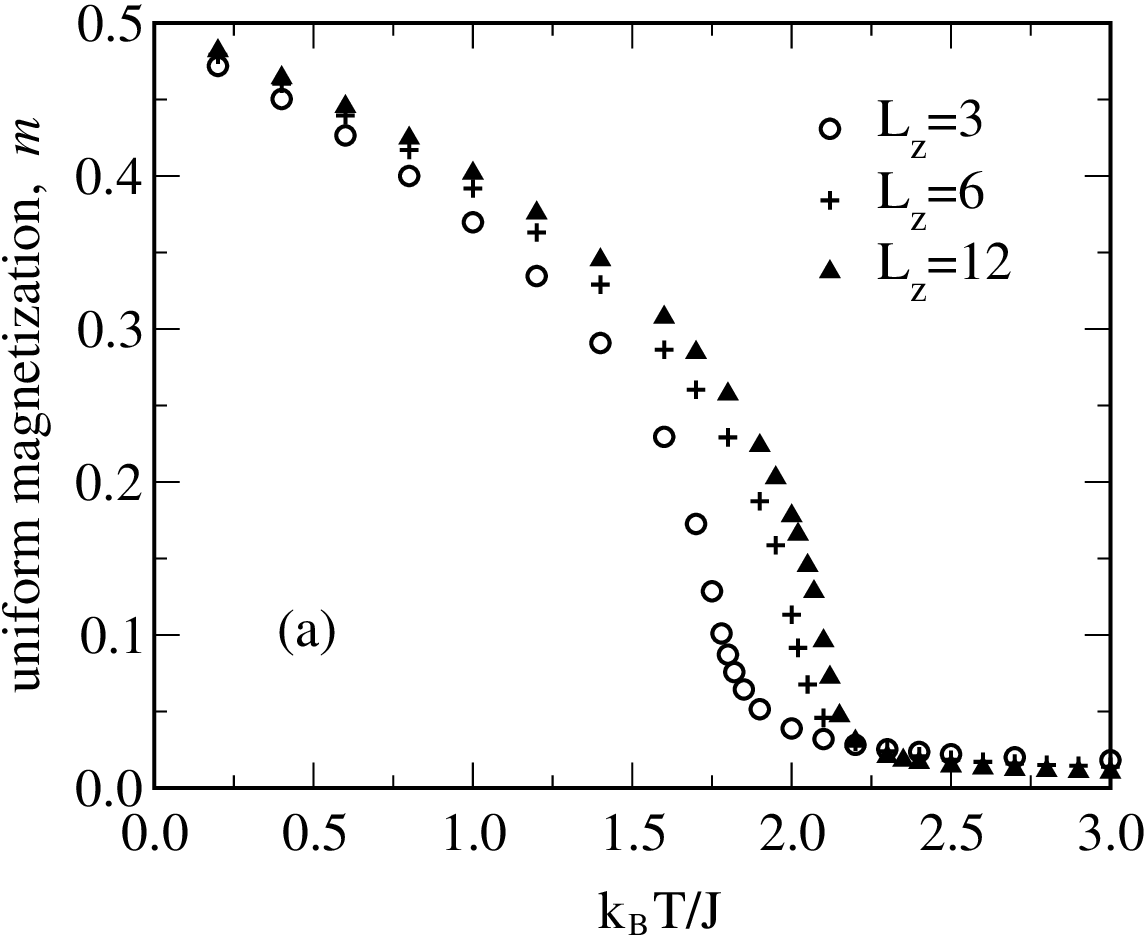}
\includegraphics[clip,angle=0,width=8cm]{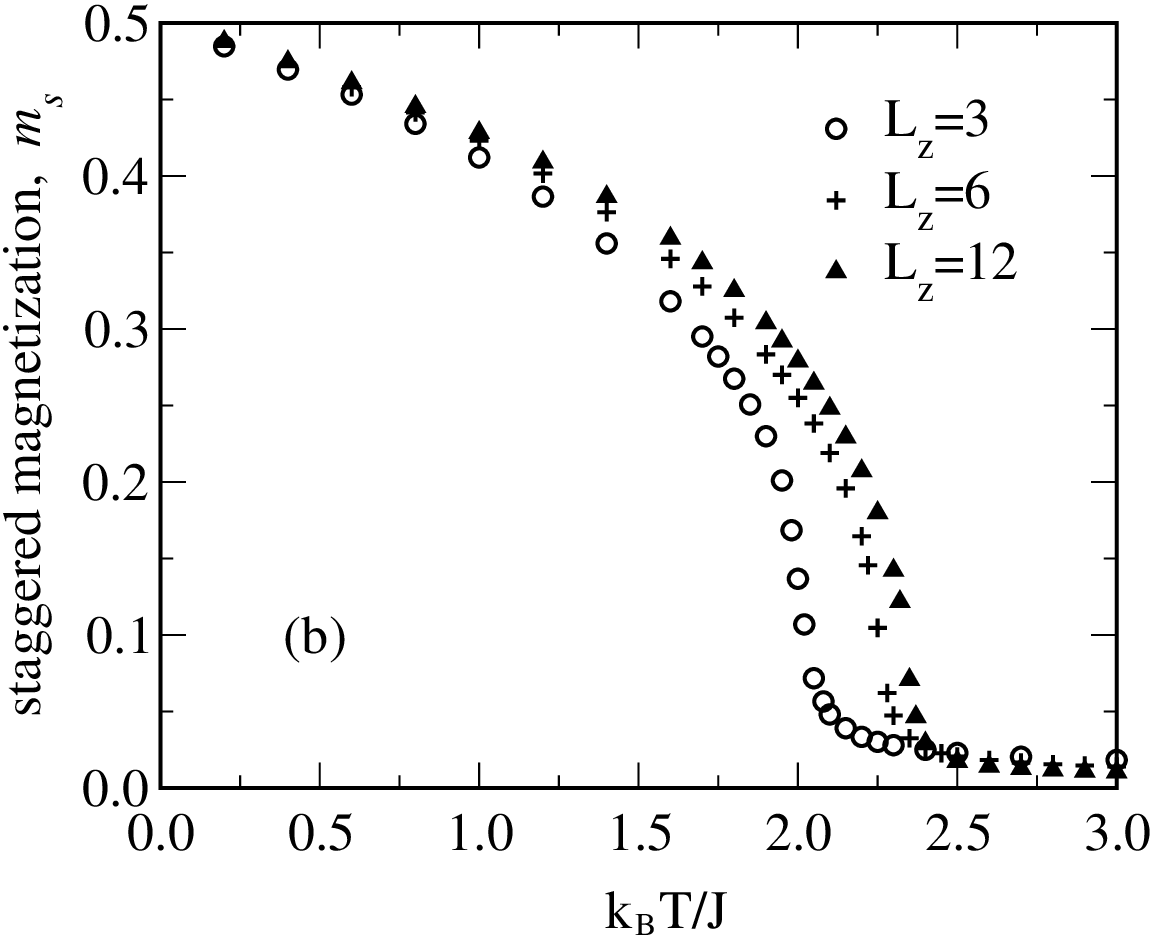}
\caption{\label{figthick}(a) Uniform and (b) staggered magnetizations as 
a function of temperature for several film thicknesses, with $L_x=L_y=20$ and 
$L_z^A=L_z^F=L_z$.}
\end{figure}

\begin{figure}[ht]
\includegraphics[clip,angle=0,width=8cm]{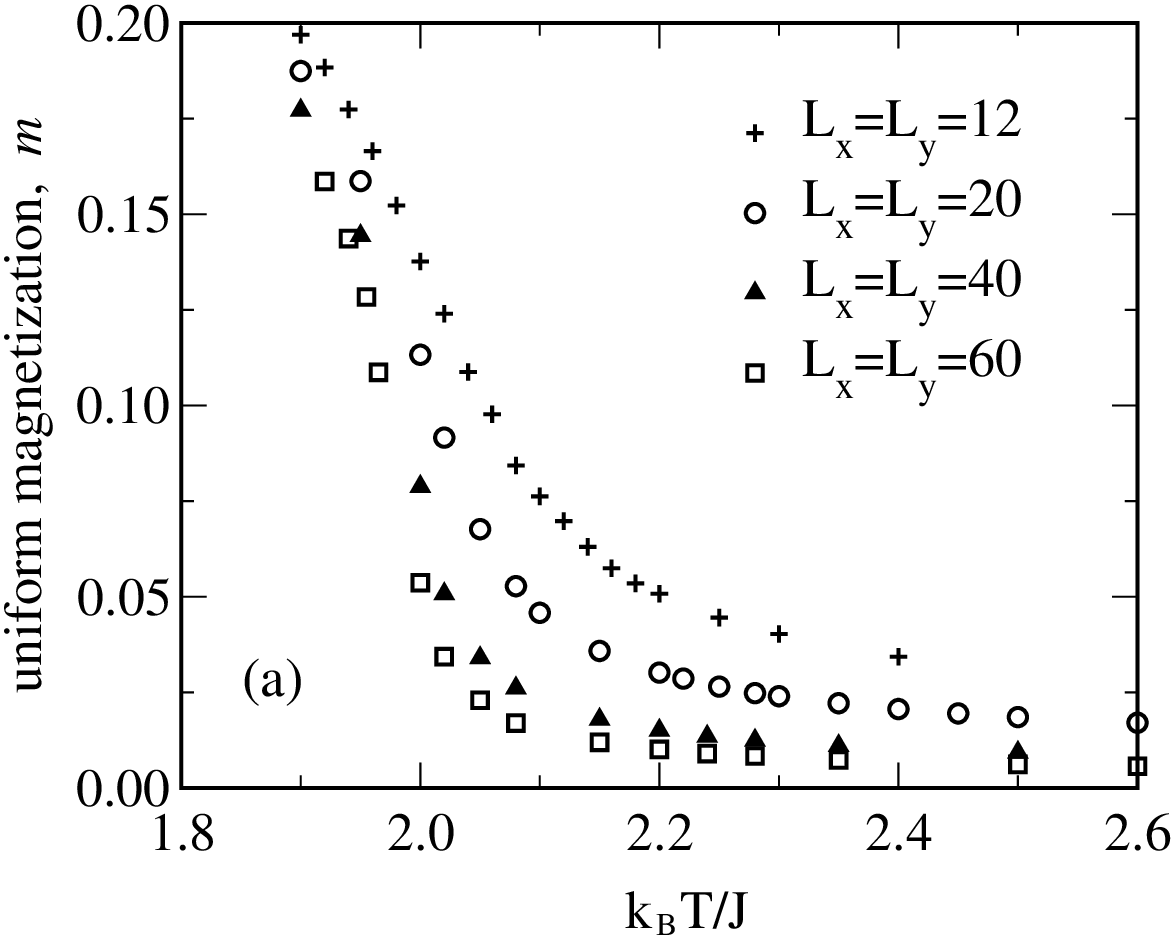}
\includegraphics[clip,angle=0,width=8cm]{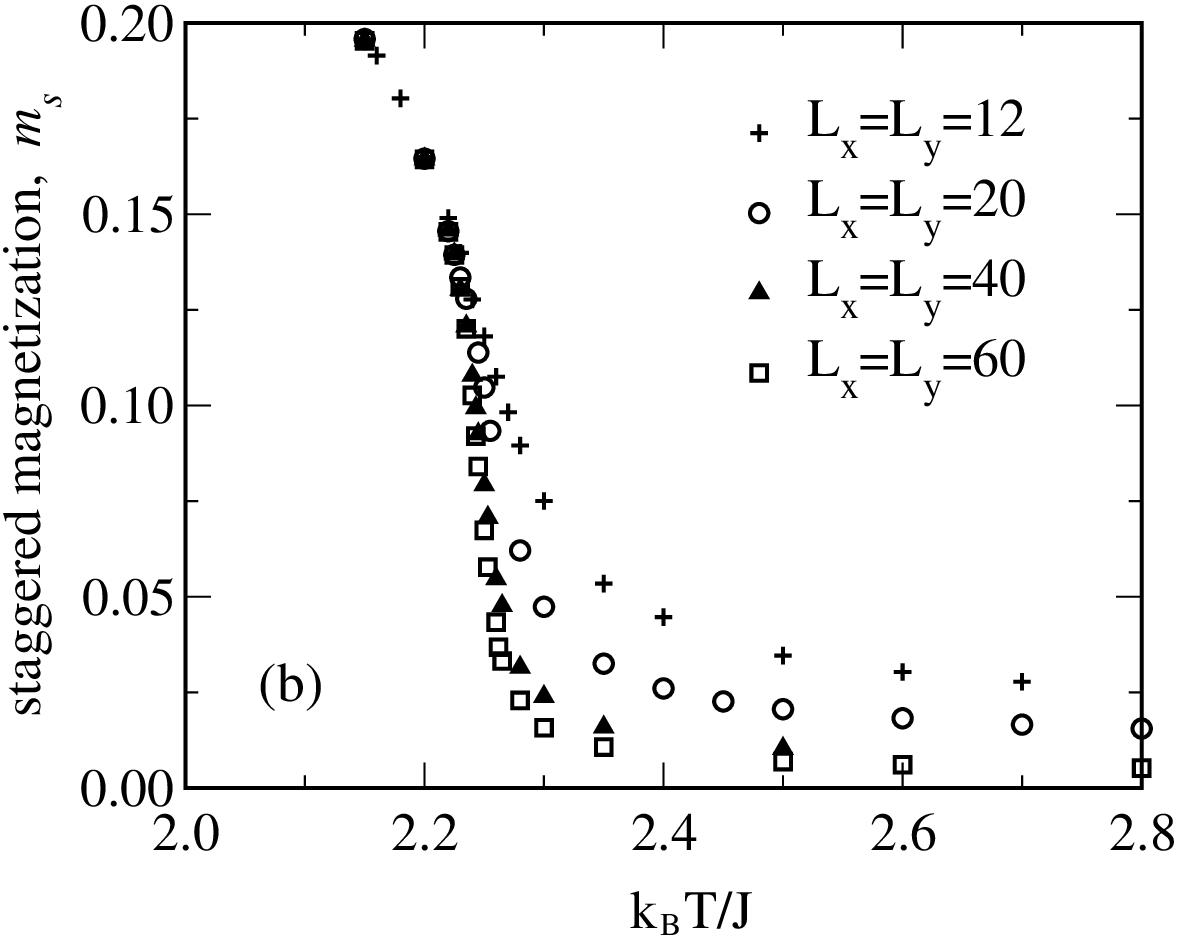}
\caption{\label{figcross}(a) Uniform and (b) staggered magnetizations as a 
function of temperature, 
for $L_z^A=L_z^F=6$ and several cross sections.}
\end{figure}

Figs.(\ref{figthick}a) and (\ref{figthick}b) show the uniform and staggered
magnetizations per spin for films of different thicknesses and fixed cross 
section.
At low temperature, both $m$ and $m_s$ tend to 0.5, indicating ordered FM and 
AFM films (note that both quantities are computed for the entire lattice and 
thus
go to 0.5 instead of 1.0 when the films are ordered). As the temperature 
increases, both $m$ and $m_s$ decay to zero, indicating disordered spin
configurations on both films at higher temperatures.
In simulations of finite lattices, occurrence of phase 
transitions cannot be ascertained with the use of only one lattice size. 
Therefore, we have considered different cross sections for each film thickness.
Illutrations of finite-size effects on the uniform and the staggered 
magnetizations near the phase transitions are presented in 
Figs.(\ref{figcross}a) and (\ref{figcross}b), respectively. These figures 
show that as the cross section increases, the decay of the magnetizations 
$m$ and $m_s$ to zero becomes sharper. Such dependence of order 
parameters on finite lattice sizes is characteristic of real phase 
transitions. Similar finite-size effects were observed for the uniform 
and the staggered magnetizations of 
films of other thicknesses. The behavior of the uniform and staggered 
magnetizations shown in Figs.(\ref{figthick}a) and (\ref{figthick}b), with
finite-size effects described above, suggests that for a given film thickness
there are two distinct phase transitions, one occuring on the AFM film and 
one on the FM film. The AFM transition temperature 
seems to be slightly higher than the FM one, presumably due to the higher 
anisotropy on the AFM film. In addition, it appears that the 
transition temperatures on both the FM and the AFM films increase with 
film thickness. 
We use crossings of reduced fourth-order cumulants of order 
parameters \cite{binder}, defined 
as $U_{4r}=1-\langle m^4\rangle/(3\langle m^2\rangle^2)$ and 
$U_{4r}^s=1-\langle m_s^4\rangle/(3\langle m_s^2\rangle^2)$, to locate 
the phase transition temperatures on the FM and AFM films, respectively. 
Preliminary analyses of cumulant crossings indicate that the AFM transition
for $L_z^A=L_z^F=6$ occurs at $T_c=(2.235\pm 0.005)J/k_B$, and it is consistent
with the two-dimensional Ising universality class. For the same film thickness,
the FM transition appears to be at $T_c=(1.95\pm 0.01)J/k_B$; however, the 
nature of this transition has not been determined yet. 

\begin{figure}[ht]
\includegraphics[clip,angle=0,width=8cm]{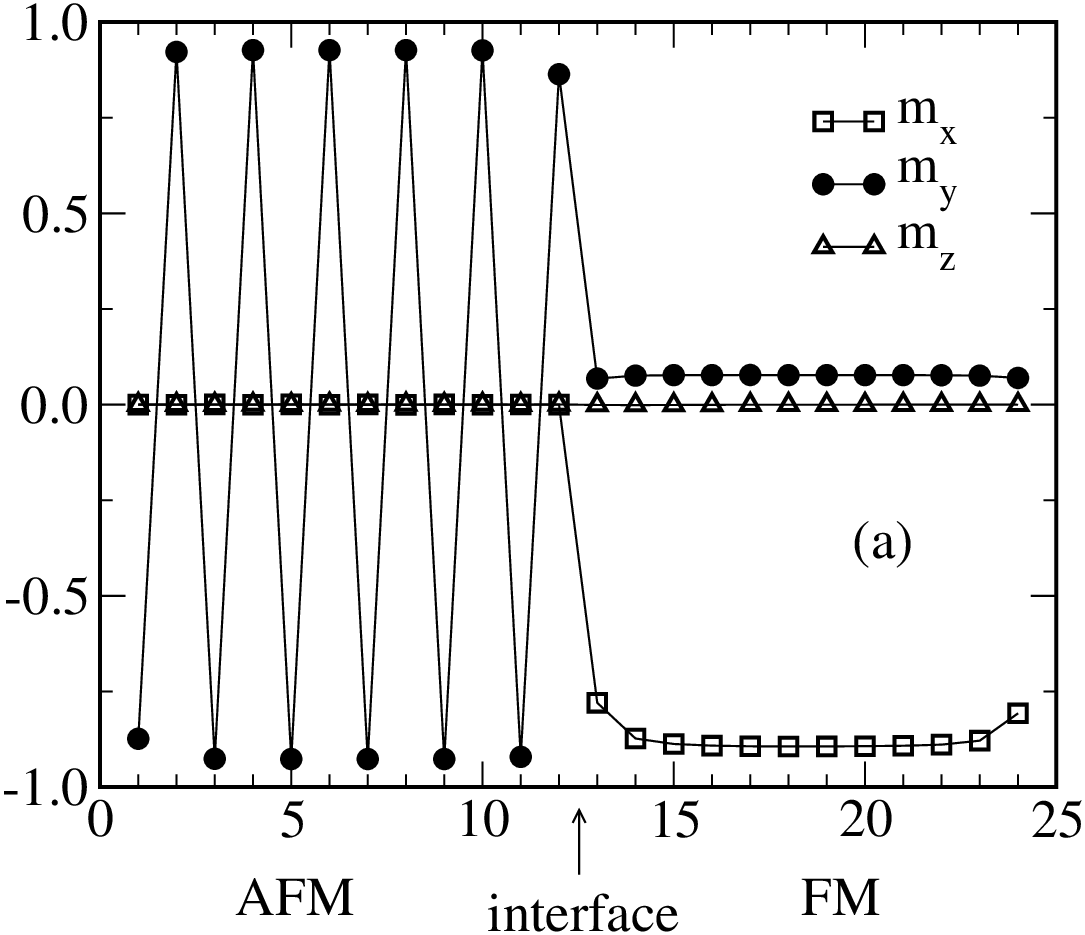}
\includegraphics[clip,angle=0,width=8cm]{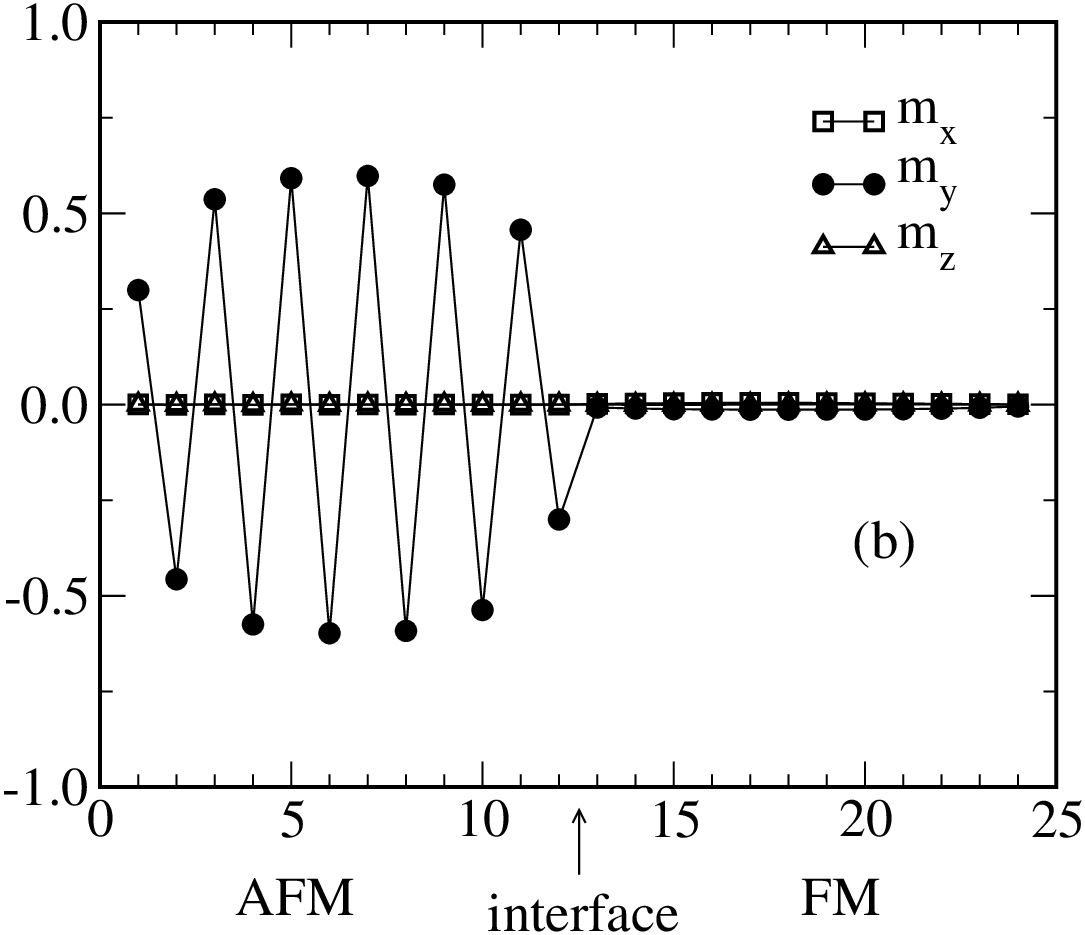}
\caption{\label{figlayer}Components of the uniform magnetization as a function
 of lattice layer, for $L_z^A=L_z^F=6$, $L_x=L_y=20$ and (a) $T=0.6J/k_B$ and
(b) $T=2.0J/k_B$. The solid lines are guides to the eyes.}
\end{figure}

Figs.(\ref{figlayer}a) and (\ref{figlayer}b) show the $x$, $y$, and $z$ 
components of the uniform magnetization per site as
a function of film layer, for $L_z^A=L_z^F=6$, at $T=0.6J/k_B$ and 
$T=2.0J/k_B$, respectively.
While Fig.(\ref{figlayer}a) illustrates the behavior of
the layer magnetizations when both films are in the respective ordered states,
Fig.(\ref{figlayer}b) corresponds to a temperature above the FM transition
and below the AFM one. In our notation, layers 1 to 12 belong to the 
AFM film, whereas layers 13 to 24 comprise the FM film. 
Layer magnetizations
on the AFM film are almost entirely along the $y$-axis, which is the easy
axis for the uniaxial anisotropy on this film. Because of the BCC
lattice structure, spins on consecutive layers belong to different 
sublattices; hence consecutive layer magnetizations on the AFM film 
have different sign. In contrast, layer magnetizations on the FM film
in the ordered state are
essentially parallel, with a large component along the $x$-axis. 
In the absence of the AFM film, spins on the FM film have global rotation
symmetry on the $x$-$y$ plane, which is the easy plane for spins on 
this film. However, coupling to the AFM film causes
spins on the FM layers to orient preferentially in a direction perpendicular
to the AFM easy axis. The slightly lower values of the
magnetizations for layers 1 and 24 in Fig.(\ref{figlayer}a) are due to the
free boundary conditions used along the $z$ direction. Similar boundary
effects are seen in Fig.(\ref{figlayer}b).

\section{Conclusions}
We have used extensive Monte Carlo simulations to study a system of 
coupled FM/AFM films, with compensated exchange across the
interface. For the values of anisotropies and interaction parameters 
considered, it appears that the FM and the AFM phase transitions occur 
at different temperatures. The latter transition is consistent with 
the two-dimensional Ising universality class. Our preliminary simulations
suggest that both phase transition temperatures increase with the 
film thickness. 

\section{Acknowledgments}
This research was partially supported by NSF grant DMR-0094422 and 
ORNL Laboratory Directed R\&D Fund as well as DOE-OS through BES-DMSE
and OASCR-MICS under subcontract No. DE-AC05-00OR22725 with
UT-Battelle. Simulations were performed on the IBM SP at NERSC.


\end{document}